\documentclass{ptephy_v1}

\preprintnumber{XXXX-XXXX} 

\usepackage{amsmath,amssymb}
\usepackage{braket}
\usepackage{bm}
\usepackage{url}

\begin{document}

\title{Development of microwave cavities for measurement of muonium hyperfine structure at J-PARC}


\author[1,4]{K.S.~Tanaka}
\author[2]{M.~Iwasaki}
\author[2]{O.~Kamigaito}
\author[3,6,7]{S.~Kanda}
\author[3,6,7]{N.~Kawamura}
\author[4]{Y.~Matsuda} 
\author[5,6,7]{T.~Mibe}
\author[3,6]{S.~Nishimura}
\author[6,8]{N.~Saito}
\author[2]{N.~Sakamoto}
\author[2,4]{S.~Seo}
\author[3,6,7]{K.~Shimomura}
\author[3,6,7]{P.~Strasser}
\author[2]{K.~Suda}
\author[2,4]{T.~Tanaka}
\author[4,8]{H.A.~Torii}
\author[5,6,7]{A.~Toyoda}
\author[2,4]{Y.~Ueno}
\author[7,9]{M.~Yoshida}

\affil[1]{Cyclotron and Radioisotope Center, Tohoku University, 6-3 Aoba, Aramaki, Aoba-ku, Sendai, Miyagi 980-8578, Japan}
\affil[2]{RIKEN, 2-1 Hirosawa, Wako, Saitama 351-0198, Japan}
\affil[3]{Institute of Materials Structure Science, KEK, 1-1 Oho, Tsukuba, Ibaraki 305-0801, Japan}
\affil[4]{Graduate School of Arts and Sciences, The University of Tokyo, 3-8-1 Komaba, Meguro-ku, Tokyo 153-8902, Japan}
\affil[5]{Institute of Particle and Nuclear Studies, KEK, 1-1 Oho, Tsukuba, Ibaraki 305-0801, Japan}
\affil[6]{Japan Proton Accelerator Research Complex (J-PARC), 2-4 Tokai, Ibaraki 319-1195, Japan}
\affil[7]{Graduate University of Advanced Studies (SOKENDAI), 1-1 Oho, Tsukuba, Ibaraki 305-0801, Japan}
\affil[8]{School of Science, The University of Tokyo, Bunkyo, Tokyo 113-0033, Japan}
\affil[9]{Accelerator Laboratory, KEK, 1-1 Oho, Tsukuba, Ibaraki 305-0801, Japan}


\begin{abstract}%
   The MuSEUM collaboration is planning measurements of the ground-state hyperfine structure (HFS) of muonium at the Japan Proton Accelerator Research Complex (J-PARC), Materials and Life Science Experimental Facility. The high-intensity beam that will soon be available at H-line allows for more precise measurements by one order of magnitude. We plan to conduct two staged measurements. First, we will measure the Mu-HFS in a near-zero magnetic field, and thereafter we will  measure it in a strong magnetic field. We have developed two microwave cavities for this purpose.
   Furthermore, we evaluated systematic uncertainties from such a fluctuation of microwave fields and confirm the requirement of the microwave system, we use a microwave field distribution calculated from the finite element method.
   
\end{abstract}

\subjectindex{C31}

\maketitle

\section{Introduction}
Muonium (Mu) is a hydrogen-like bound state that consists only of leptons, and its ground-state hyperfine structure (HFS) provides an effective probe for testing the bound-state quantum electrodynamics (QED) theory. Furthermore, it can be used to extract the muon--proton magnetic moment ratio 
($\mu_{\mu} / \mu_{p} $) and muon--electron mass ratio ($m_{\mu} / m_{e} $). This has been indispensable for determining the value of the muon anomalous magnetic moment ($g$--2) in measurements at the Brookhaven National Laboratory (BNL) \cite{PhysRevD.73.072003}, and the discrepancy between the results and the theoretical prediction values \cite{PhysRevD.98.030001} has been hotly debated. The measured transition frequencies are sensitive to the sidereal oscillation that would originate from additional Hamiltonian terms, which violate CPT and Lorentz invariance \cite{PhysRevLett.87.111804}.

\par
The experimental value of the Mu-HFS in CODATA is $4.463 302 765(53)\ \mathrm{GHz}\ (12 \ \mathrm{ppb})$, which has primarily been determined by measurements at the Los Alamos Meson Physics Facility (LAMPF) \cite{Liu_1999}.
This value was measured in a strong magnetic field.
In a magnetic field, the Zeeman effect splits the energy levels of the muonium ground-state. The spin Hamiltonian of muonium is expressed as \cite{hughes1977muon}
\begin{equation}
\mathcal{H}=h\nu_{\mathrm{HFS}} \bm{S_{\mu}} \cdot \bm{S_{e}} + g'_{e}\mu^{B}_{e} \bm{S_{e}}\cdot \bm{B} - g'_{\mu} \mu^{B}_{\mu}\bm{S_{\mu}} \cdot \bm{B},
\label{hamiltonian-equation}
\end{equation}
where $h\nu_{\mathrm{HFS}}$ is the HFS coupling constant, $\bm{S_{\mu}}$ is the muon spin operator, $\bm{S_{e}}$ is the electron total angular momentum operator, $\bm{B}$ is the external static magnetic field, and $\mu_{e}$ and $\mu_{\mu}$ are the electron and muon Bohr magnetrons. Moreover, $g'_{e}$ and $g'_{\mu}$ are the g-factors of an electron bound in muonium and of a muon in muonium.
Under zero external magnetic fields ($B = \bm{0}$), only the first HFS term remains in Eq. (\ref{hamiltonian-equation}), so that the HFS can be derived directly with spectroscopy.
The result of the most recent zero-field experiment\cite{1975PhLB...59..397C} was $4. 463 302 2(14)  \  \mathrm{GHz}\ (0.3 \ \mathrm{ppm})$. 
One of the systematic uncertainties of the zero-field experiment arises from the strength of the magnetic field, which was suppressed at less than 100 nT. 
\par
\begin{figure}[tbp]
\begin{center}
   \includegraphics[width=\hsize,keepaspectratio]{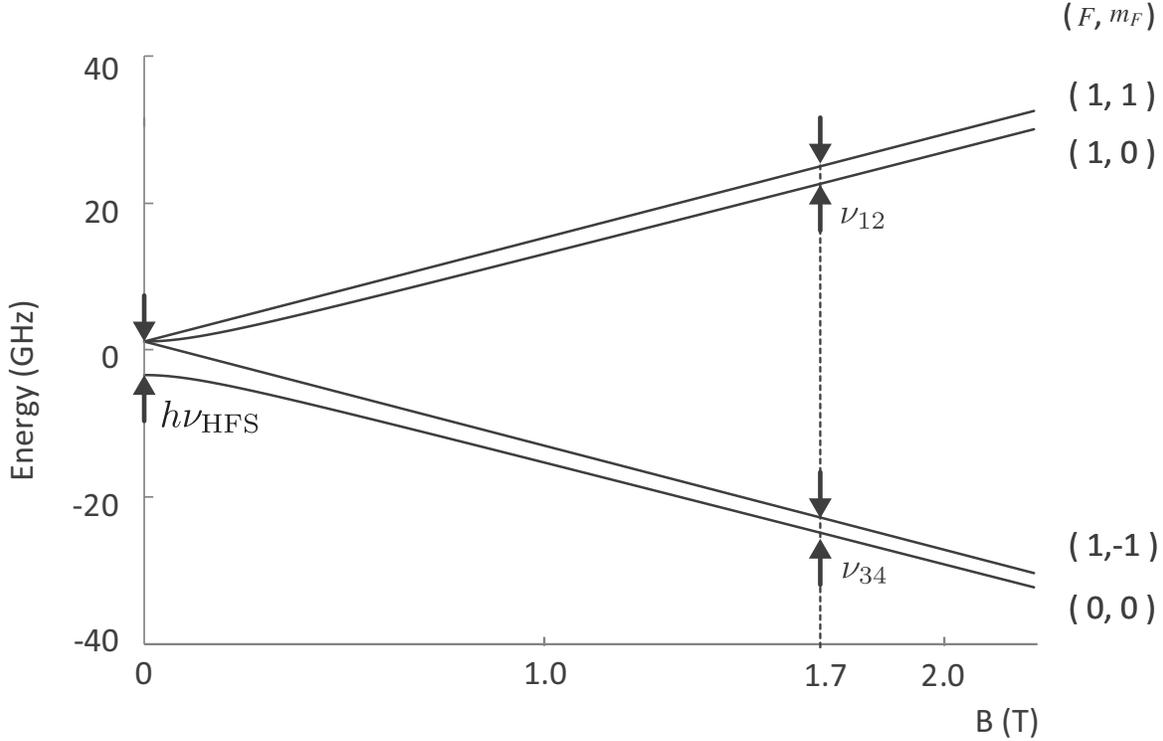}
   \caption{Energy level diagram of muonium as function of external magnetic field. The Mu-HFS ($h\nu_{\mathrm{HFS}}$) is obtained by $\nu_{12}$ and $\nu_{34}$ in a common strong magnetic field.}
   \label{breit-rabi}
\end{center}
\end{figure}

Under the strong external magnetic field ($\bm{B} > \bm{0} $),
eigenvalues of the Hamiltonian presented in Eq. (\ref{hamiltonian-equation}), can be organized as
\begin{align}
E_{F=1,M_{F}=1} &=& +&\frac{1}{4}h\nu_{\mathrm{HFS}} + \frac{1}{2}h\nu_{\mathrm{HFS}}(g'_{e}\mu_{e}^{B} -g'_{\mu} \mu^{B}_{\mu})B \label{breit-rabi-formula1}, \\
E_{F=1,M_{F}=0} &=& -&\frac{1}{4}h\nu_{\mathrm{HFS}} + \frac{1}{2}h\nu_{\mathrm{HFS}}\sqrt{1+x^2}\label{breit-rabi-formula2},\\
E_{F=1,M_{F}=-1} &=& +&\frac{1}{4}h\nu_{\mathrm{HFS}} -\frac{1}{2}h\nu_{\mathrm{HFS}}(g'_{e}\mu_{e}^{B} -g'_{\mu} \mu^{B}_{\mu})B\label{breit-rabi-formula3}, \\
E_{F=0,M_{F}=0} &=& -&\frac{1}{4}h\nu_{\mathrm{HFS}} - \frac{1}{2}h\nu_{\mathrm{HFS}}\sqrt{1+x^2}\label{breit-rabi-formula4},
\end{align}
where $F$ is the quantum number of the total angular momentum, $M_{F}$ is the magnetic quantum number of the total angular momentum, and $x$ is a parameter that is proportional to the magnetic field ($B > \bm{0}$):
\begin{equation}
x = \frac{g'_{e}\mu^{e}_{B}+g'_{\mu}\mu^{B}}{h\nu_{\mathrm{HFS}}} B.
\end{equation}
Figure \ref{breit-rabi} displays the Breit--Rabi diagram, which indicates that the ground-state splits into four substates in a static magnetic field. The transitions for ($F = 1, M_{F} = 0$) $\leftrightarrow$ ($F = 1,M_{F} = 1$) are designated as $\nu_{12}$ and those for ($F = 0, M_{F} = 0$) $\leftrightarrow$ ($F = 1, M_{F} = -1$) are designated as $\nu_{34}$. These transition energies were calculated from Eqs. (\ref{breit-rabi-formula1}) to (\ref{breit-rabi-formula4}), as follows:
\begin{align}
h\nu_{12} &=&\ \  -&g'_{\mu}\mu_{\mu}^{B}B + \frac{h\nu_{\mathrm{HFS}}}{2}[1+x-\sqrt{1+x^{2}}] \label{nu12-formula},\\
h\nu_{34} &=& &g'_{\mu}\mu_{\mu}^{B}B + \frac{h\nu_{\mathrm{HFS}}}{2}[1-x+\sqrt{1+x^{2}}] \label{nu34-formula}.
\end{align}
According to Eqs. (\ref{nu12-formula}) and (\ref{nu34-formula}), the Mu-HFS can be expressed as
\begin{equation}
   \nu_{\mathrm{HFS}}= \nu_{12}+\nu_{34}. \label{nu12-nu34}
\end{equation}
Thus, the Mu-HFS transition energy is obtained by summing $\nu_{12}$ and $\nu_{34}$ in a common strong magnetic field, and it is independent of the absolute value of the magnetic field.
Further motivation for measurement in a high magnetic field to provide the fundamental constants of the muon, such as the muon--proton magnetic moment ratio ($\mu_{\mu}/\mu_{\mathrm{p}}$) and muon--electron mass ratio ($m_{\mu}/m_{\mathrm{e}}$) \cite{doi:10.1063/1.4954402}. Thus far, these have been determined by HFS experiments as
\begin{eqnarray}
\frac{\mu_{\mu}}{\mu_{\mathrm{p}}} &=& 3.183 345 24(37)\ (120\ \mathrm{ppb}),  \\
\frac{m_{\mu}}{m_{\mathrm{e}}} &=& 206.768 276 (24)\ (120\ \mathrm{ppb}).
\end{eqnarray}

The precise determinations of $\mu_{\mu}/\mu_{\mathrm{p}}$ and $m_{\mu}/m_{p}$ by means of Mu-HFS spectroscopy can contribute to understanding the muon anomalous magnetic moment (muon $g$--2). The BNL E821 experiment measured the muon and a 3.5 $\sigma$ deviation from the calculation was reported based on the Standard Model \cite{PhysRevD.73.072003}. A new $g$--2 measurement is being planned at the Japan Proton Accelerator Research Complex (J-PARC) \cite{10.1093/ptep/ptz030} and Fermi National Accelerator Laboratory (FNAL) \cite{refId0}.
These experiments will measure the difference $\omega_{a}$ between the Larmor angular frequency and the cyclotron angular frequency of the muon, following which the anomalous magnetic dipole moment ($a_{\mu} = (g-2)/2$) can be expressed as
\begin{equation}
\frac{\omega_{a}}{\omega_{\mathrm{p}}} = \frac{a_{\mu}}{1+a_{\mu}} \frac{\mu_{\mu}}{\mu_{\mathrm{p}}},
\end{equation}

where $\omega_{p}$ is the Larmor precession angular velocity of a proton in water.
Thus, the precision of the anomalous magnetic dipole moment from the $g$--2 experiment depends on that of $\mu_{\mu}/\mu_{\mathrm{p}}$.
The targeted precisions at both the J-PARC and FNAL are at a level of 100 ppb, which is comparable to that of $\mu_{\mu}/\mu_{\mathrm{p}}$, so that updating of the values will be required for these projects.
\par
In the first phase of our project, we plan to measure the Mu-HFS in a zero magnetic field and to achieve the same order of precision as the most recent zero-field experiment. Thereafter, we will shift the experiment to a strong magnetic field with similar apparatus. We aim to achieve determinations of Mu-HFS, $\mu_{\mu}/\mu_{\mathrm{p}}$, and $m_{\mu}/m_{\mathrm{e}}$ that are one order of magnitude more precise.
\par
The precision of the latest high-field experiment result has mainly been limited by statistics. However, there is a reason that there no new measurements have been performed to improve the statistics.
The muon beam that was available at the LAMPF is a continuous beam, which means that the muons are intermittently delivered one by one \cite{THOMPSON1979391}\cite{REIST197861}. When we wish to observe the transition of muonium atoms under a microwave field, we need to chop the muon beam to provide it with time to arrive, which causes the effective beam intensity to be drastically reduced. For example, the most recent Mu-HFS measurement at the LAMPF chopped the muon beam to 3.9 $\mu$s pulses separated by 9.9 $\mu$s when a microwave field was applied \cite{Liu-phd-thesis}. This means that, although the LAMPF accelerator can provide approximately 1$\times$10$^7$ muons/s, the experiment used less than 30 \% of the available muons for the measurement.
\par
The inauguration of J-PARC Materials and Life Science Experimental Facility (MLF) and its H-line transformed the entire picture. The J-PARC MLF provides a double-pulsed muon beam (600 ns intervals in 25 Hz repetitions), which means that all muons are bunched into pulses, and their arrival time to the target can be determined with a high-rate capable particle detector \cite{Pos2015kanda} for decay positrons. Therefore, our experiment at J-PARC can make full use of the available muons, thereby offering a significant advantage over previous experiments. Furthermore, the capture solenoid on the H-line, which provides surface muons for fundamental physics experiments, enables high transmission efficiency of more than 80 \% of the captured muons, and the beam intensity is estimated as $10^{8}$/s at 1 MW beam power \cite{10.1093/ptep/ptz030}. Under these conditions, we can achieve two orders of magnitude better statistics with a 100-day measurement than those of the 6-week measurement of LAMPF experiment.
\section{Experimental procedure}
\begin{figure}[tbp]
\begin{center}
\includegraphics[width=\hsize]{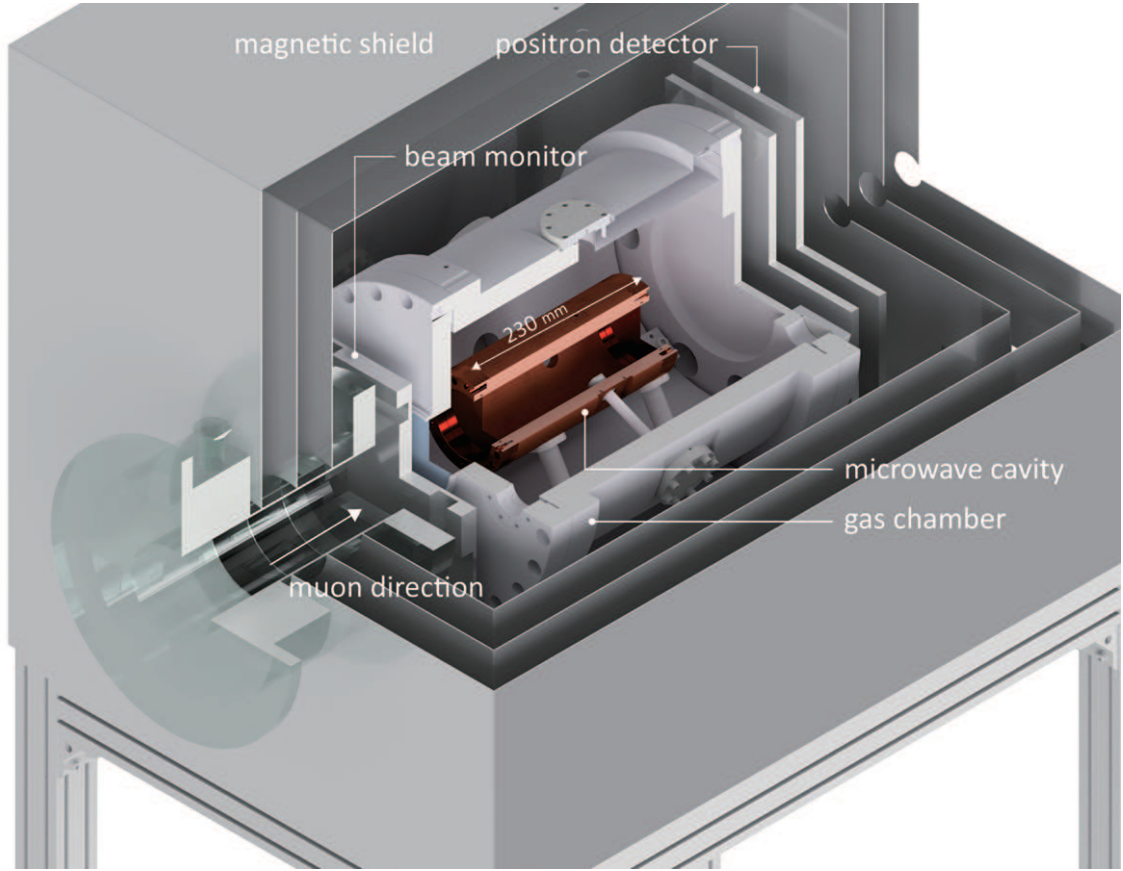}
\caption{Schematic of experimental setup in zero-field.}
\label{drawing-of-overview-zerofield}
\end{center}
\end{figure}

\begin{figure}[tbp]
\begin{center}
\includegraphics[width=\hsize]{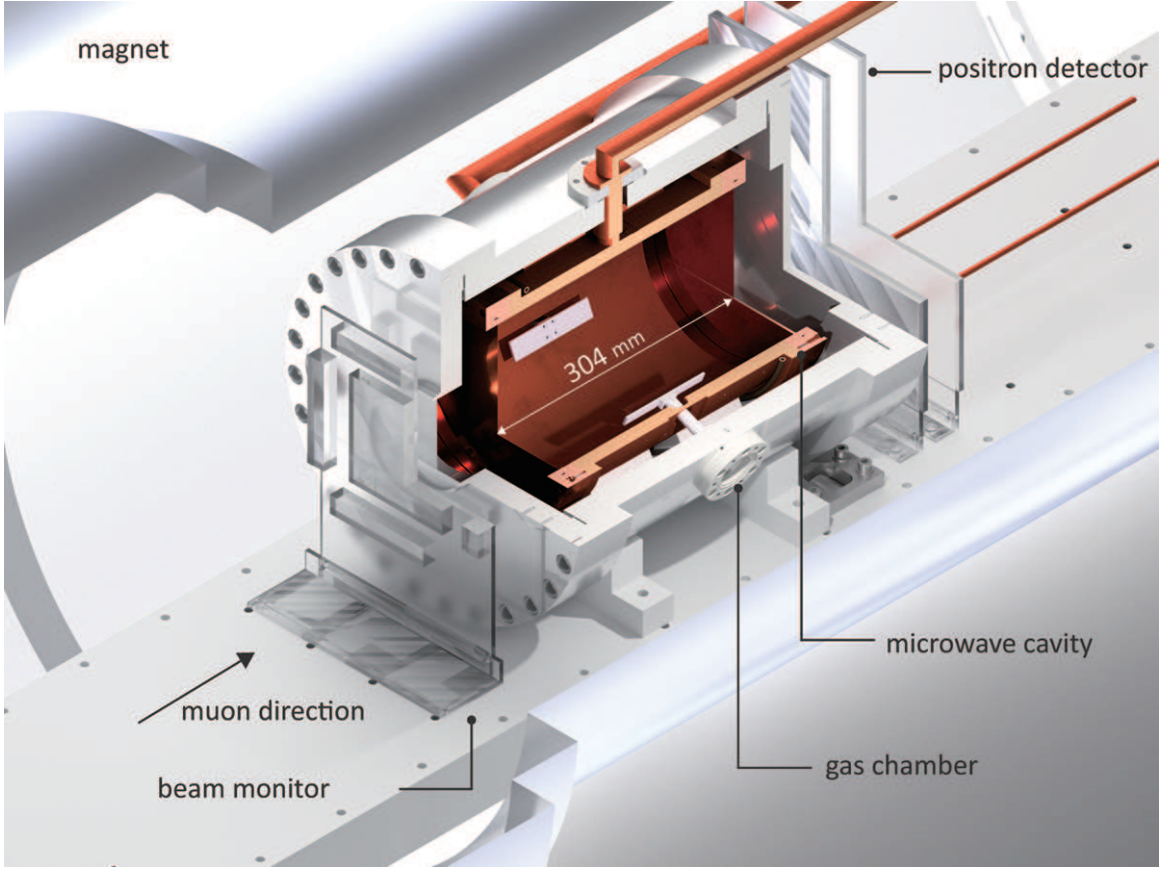}
\caption{Schematic of experimental setup in high-field.}
\label{drawing-of-overview-highfield}
\end{center}
\end{figure}

The measurement procedures for both the high-field and the zero-field experiments are introduced as follows (Figs. \ref{drawing-of-overview-zerofield} and \ref{drawing-of-overview-highfield}).
\begin{enumerate}
\item The intense polarized pulsed muon beam is provided by the J-PARC MLF muon beamline.
\item The microwave cavity, which is located in a gas chamber containing pure krypton (Kr) gas, is either inserted into a large superconducting solenoid for the high-field measurement or enclosed in a magnetic shield for the zero-field measurement. The muons are stopped by collisions in the gas and polarized muonium is formed by the electron capture process with the Kr.
\item Decay positrons are emitted preferentially in the direction of the muon spin. By driving the transitions with a microwave magnetic field, the muon spin can be flipped and the angular distribution of high momentum positrons changes from predominantly upstream to downstream with respect to the beam direction.
\end{enumerate}
During microwave frequency sweep, the cavity resonance frequency is changed by the displacement of the tuning bars in the cavity. Moreover, the microwave power and distribution of the microwave field in the cavity are varied. If sufficient events are accumulated, the fluctuations in the microwave power and distribution of the microwave field in the cavities will form a serious part of the systematic uncertainties; thus, the design of the microwave system is crucial to ensuring the highest precision measurement. We have developed two microwave cavities and a microwave circuit for the high-field and zero-field experiments. Both systems possess sufficient tunability to carry out a scan of the transition frequencies, and a power feedback system to suppress the power fluctuation within 1 \%. The systematic uncertainties have been estimated by a full Monte-Carlo simulation, and satisfy the conditions to achieve a precision of several ppb in the measurement.
\par
The most dominant part of the uncertainties of $\nu_{\mathrm{Mu}}$ in the LAMPF experiment was a gas pressure shift, which can be improved by a longer length of the cavity. The estimation of the uncertainty from the gas pressure is discussed in subsection \ref{subsection-cavity-length}.
\section{Microwave cavity}
This section describes the requirements and design of the microwave cavity.
\par
As the muon decays from the outside of the cavity causing background noise, the cavity should be larger than the muon stopping distribution.
\par
Furthermore, the preferred microwave field is the magnetic field strength of microwave perpendicular
to the cavity axis maximized at the center of the cavity, which is the region in which muonium is formed, for a higher transition probability. Replicability of the microwave intensity distribution is also important to suppress the uncertainty from the muonium distribution.
\subsection{Resonance modes and cavity diameter}
\begin{figure}[htbp]
  \begin{center}
     \includegraphics[width=0.7\hsize,keepaspectratio]{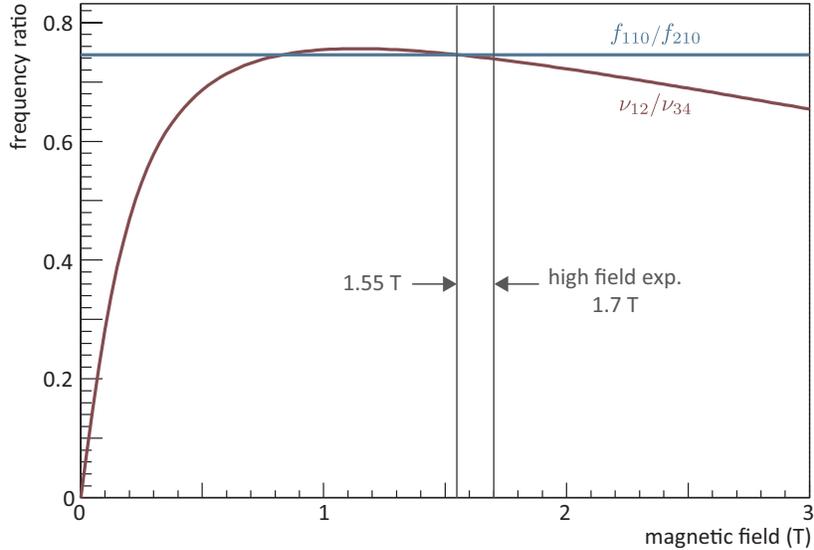}
     \caption{Dependence of $\nu_{12}/\nu_{34}$ on magnetic field. As the ratio of the TM110 and TM210 modes is constant for a cylindrical cavity, there are only two solutions to be tuned in for both transitions. Since the relative uniformity of the magnetic field is better, higher solution which is approximately 1.55 T is preferred to the measurement, and we apply 1.7 T taking into account the effect of the tuning bar.}
     \label{resonant-mode-for-nu12-nu34}
  \end{center}
  \end{figure}

The transverse magnetic (TM) resonance modes of the cylindrical cavity are expressed as TM$mnp$, where $mnp$ denotes the number of half waves of the electric or magnetic field along the diameter, circumference, and length of the resonator. We select the TM110 mode because it does not have nodes in the axial direction and is the lowest mode in this condition. A lower mode is preferred since the diameter of the cavity can be smaller to be installed in the magnet, the microwave power density can be higher, and the  microwave power distribution is uniform. TM210 is the next higher mode and the ratio of the TM110 and TM210 frequencies corresponds to $\nu_{12}/\nu_{34}$  on the static magnetic field (Fig. \ref{resonant-mode-for-nu12-nu34}).
\par
The resonance frequency of TM$mnp$ is
\begin{equation}
  f_{mnp}=\frac{c}{n}\sqrt{(\frac{x_{mn}}{\pi D})^2+(\frac{p}{2L})^2}. \label{equation-resonance-frequency}
  \end{equation}
  In the above, $c$ is the speed of light, $D$ and $L$ are the cavity diameter and length, respectively, $n$ is the refraction index for the medium inside the cavity, and $x_{mn}$ is the $n$-th root of the Bessel function $J_{m}(x)$. 
We set the cavity diameter to 81.8 mm to adjust the resonance frequency of TM110 to the transition frequency of the zero-field experiment, and tune the frequency precisely using the tuning bar (as described in subsection \ref{subsection-cavity-structure}).
\par 
However, the two resonance frequencies of the microwave cavity for the high-field experiment should be adjusted to both the $\nu_{12}$ and $\nu_{34}$ transition frequencies.
We select TM110 and TM210, with two azimuthal nodes of the magnetic field in the resonator. The requirement for adjusting both frequencies is
\begin{equation}
  \frac{\nu_{12}}{\nu_{34}}=\frac{f_{110}}{f_{210}},
  \end{equation}
  where $f_{110}$ and $f_{210}$ are the resonance frequencies of TM110 and TM210, respectively.
As illustrated in Fig. \ref{resonant-mode-for-nu12-nu34}, the magnetic field of 1.55 T fulfills this condition. We select 1.7 T by taking into account the frequency decrease by the tuning bar made of the dielectric material.
The transition frequencies in the magnetic field of 1.7 T are $\nu_{12} \approx 1.906\ \mathrm{GHz}$ and $\nu_{34} \approx 2.556\ \mathrm{GHz}$, and the cavity diameter is 187 mm.

\subsection{Cavity length and gas pressure}\label{subsection-cavity-length}
The cavity length should be longer than the axial length of the muon stopping distribution, and the distribution depends on the Kr gas pressure.
\subsubsection{Gas pressure}
\begin{figure}[tbp]
  \begin{center}
     \includegraphics[width=\hsize,keepaspectratio]{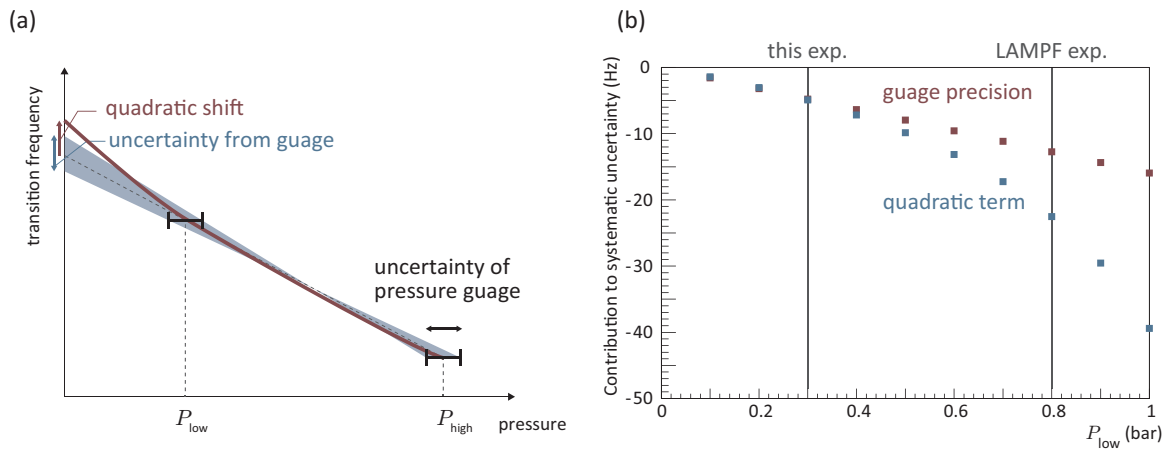}
     \caption{ (a) Transition frequencies in vacuum determined by extrapolation
     with measured frequencies at two gas pressures. (b) Dependence of systematic uncertainty from quadratic term on $P_{\mathrm{low}}$. The gas pressure is measured by a capacitance gauge (CANON ANELVA M-342DG-13), which has a measuring range of 1.33 $\times 10^{-2} $ bar with an uncertainty of $0.2\ \%$ in the reading.
     The systematic uncertainty from the quadratic term is improved to approximately the same as the systematic uncertainty from the gauge precision in the case of $P_{\mathrm{low}} = 0.8$ bar.}
     \label{gas-pressure}
  \end{center}
  \end{figure}

The transition frequencies are changed by the distortion of the wave function of muonium through collisions with the Kr atoms, and the transition frequencies in a vacuum can be extrapolated from the values at certain gas pressures $P_{\mathrm{low}}$ and $P_{\mathrm{high}}$ (Fig. \ref{gas-pressure} (a)).
As the collision rate is proportional to the Kr gas density, the shifts in the transition frequencies can be expressed as
\begin{equation}
\Delta\nu_{\rm{Kr}} = \nu_{0}(1+a_{1}P+a_{2}P^{2}),
\end{equation}
where $\nu_{0}$ is the transition frequency in a vacuum. The uncertainties from the gas pressure arise from both the quadratic term ($a_{2}P^{2}$) and the accuracy of the pressure gauge.
The quadratic coefficient $a_{2}$ was determined in a previous experiment \cite{Casperson1975397} as
\begin{equation}
   a_{2} = (5.5 \pm 1.1) \time 10^{-9}\ \mathrm{bar^{-2}}.
\end{equation}
This value was obtained by fitting the data points from two measurements at the LAMPF \cite{Casperson1975397} and Nevis \cite{PhysRevA.8.86}. The pressure range at the LAMPF was 1.7 to 5.3 bar, whereas that at Nevis was 6 to 73 bar. The effect of such a large uncertainty of $a_{2}$ can be reduced by selecting a low $P_{\mathrm{low}}$.
In the LAMPF experiment, the cavity length was 159.73 mm \cite{Liu-phd-thesis} and the gas pressure was limited by the condition that the muon stopping distribution fell within the cavity. The frequencies were measured at $P_{\mathrm{low}} = 0.8$ bar and $P_{\mathrm{high}} = 1.5$ bar to determine the transition frequencies in a vacuum by extrapolation. In the LAMPF experiment, the systematic uncertainty from the quadratic term was 1.9 ppb \cite{Liu_1999}, which is not negligible if we improve the statistical uncertainty of 12 ppb. 
\par

\begin{figure}[htbp]
   \begin{center}
           \includegraphics[width=\hsize,keepaspectratio]{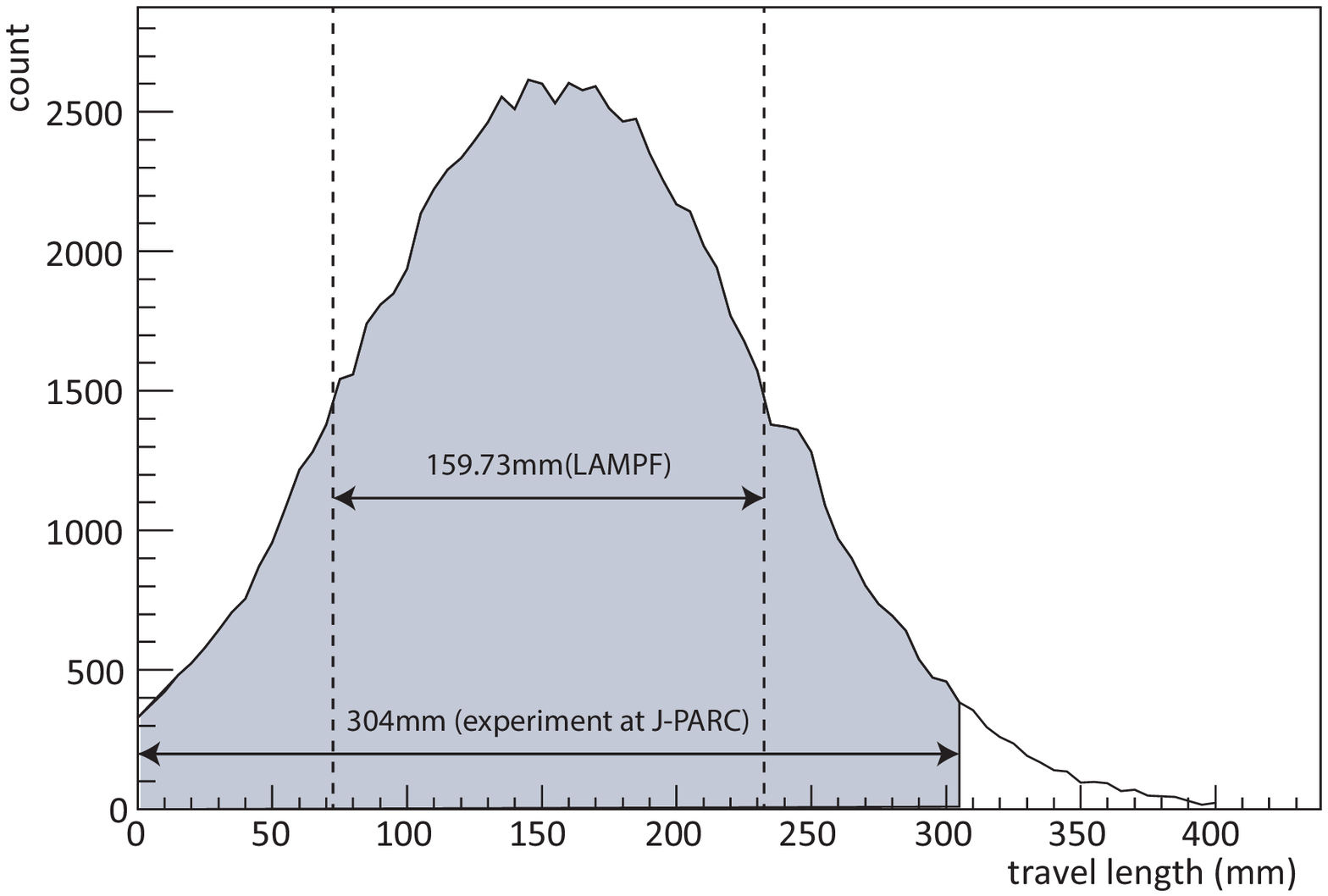}
         \caption{Muon stopping distribution with 0.3 bar Kr gas.
         The trajectories of 100000 muons are simulated using the SRIM. A total of 94 \% of muons stops within the cavity in the case of the cavity axial length of 304 mm designed for the experiment at J-PARC, whereas 68 \% of muons stops within the cavity with an axial length of 159.73 mm designed for the latest experiment at the LAMPF.}
         \label{muon-stopping-distribution-at-J-PARC}
      \end{center}
\end{figure}

Therefore, we optimize the cavity length for measurement at lower gas pressures. To suppress the systematic uncertainty from the magnetic field, the uniformity of the magnetic field in the muon stopping distribution should be considered. The magnetic field applied by a superconducting magnet has a 1 ppm uniformity in a spheroid shape with a diameter of 20 cm and an axial length of 30 cm \cite{Sasaki_2013}. This means that the maximum cavity length considering the uniformity of the magnetic field is 30 cm. According to the simulation work using the Monte-Carlo simulation program SRIM \cite{ZIEGLER20101818}, 94 \% of muons are stopped in the cavity at a gas pressure of 0.3 bar, as opposed to only 68 \% of muons that were stopped in the cavity used for the previous LAMPF experiment (Fig. \ref{muon-stopping-distribution-at-J-PARC}). The systematic uncertainty from the quadratic term will be improved five times compared to the case of $P_{\mathrm{low}} = 0.8$ bar.
   
\subsubsection{Interferences from other resonance modes}
\begin{figure}[htbp]
  \begin{center}
     \includegraphics[width=0.7\hsize,keepaspectratio]{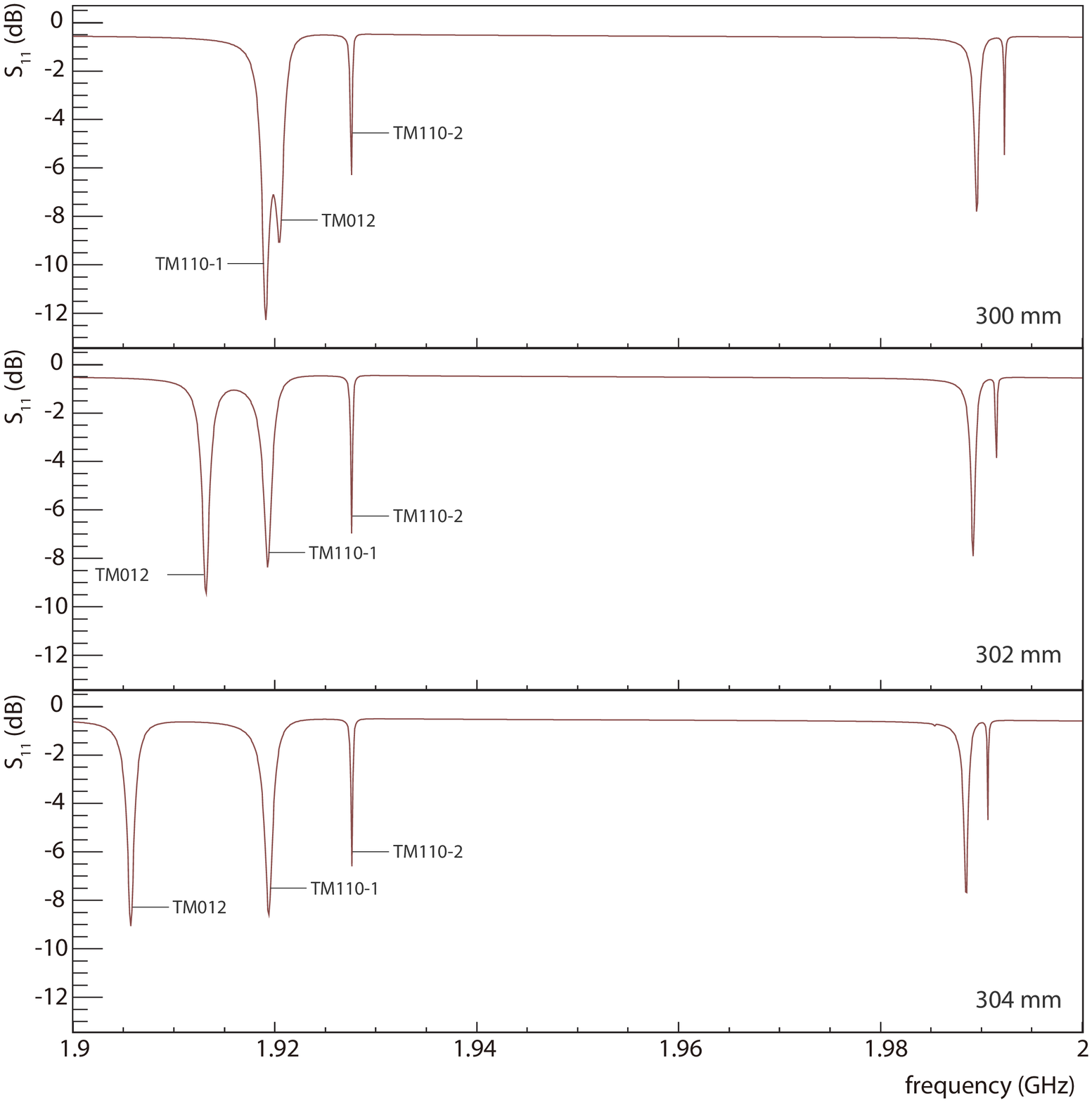}
     \caption{The simulation result of frequency characteristics of cavity with length of 300, 302, and 304 mm calculated by MWS. The TM110 mode is degenerated to TM110-1 and TM210-2 owing to the effects of the tuning bars in the cavity. Moreover, TM012 overlaps with TM110 in the case of a cavity length of 300 mm and 302 mm.}
     \label{TM110-mode-cavity-length.eps}
  \end{center}
  \end{figure}

The cavity length is also related to the condition of interferences from other resonance modes, the frequencies of which depend on the axial lengths to TM110 and TM210.
Figure \ref{TM110-mode-cavity-length.eps} represents the frequency characteristics of the cavity for the high-field experiment which is filled with a 1-atm Kr gas with cavity lengths of 300, 302, and 304 mm calculated by CST Microwave Studio (MWS) \cite{cst}. The reflective index of Kr gas (1 atm) is $n \approx 1.000429$ \cite{LEONARD197421} and changes linearly with the gas pressure. The cavity length should be 304 mm because TM012 overlaps with TM110 in the case of a cavity length of 300 mm and 302 mm. 
\par
\begin{figure}[t]
  \begin{center}
     \includegraphics[width=0.7\hsize,keepaspectratio]{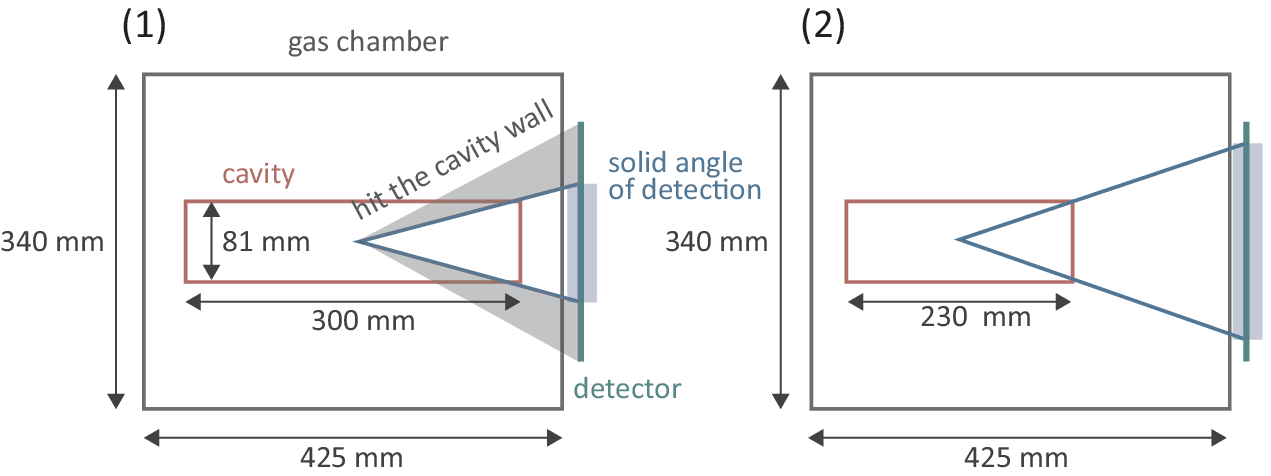}
     \caption{Schematic of the detector's position acceptance. (1) When the axial length is 300 mm, parts of the positrons hit the cavity wall before the detector (240 mm $\times$ 240 mm). (2) An axial length of 230 mm is optimal in terms of the solid angle.}
     \label{axial-length-solid-angle}
  \end{center}
  \end{figure}
\begin{figure}[htbp]
  \begin{center}
     \includegraphics[width=0.7\hsize,keepaspectratio]{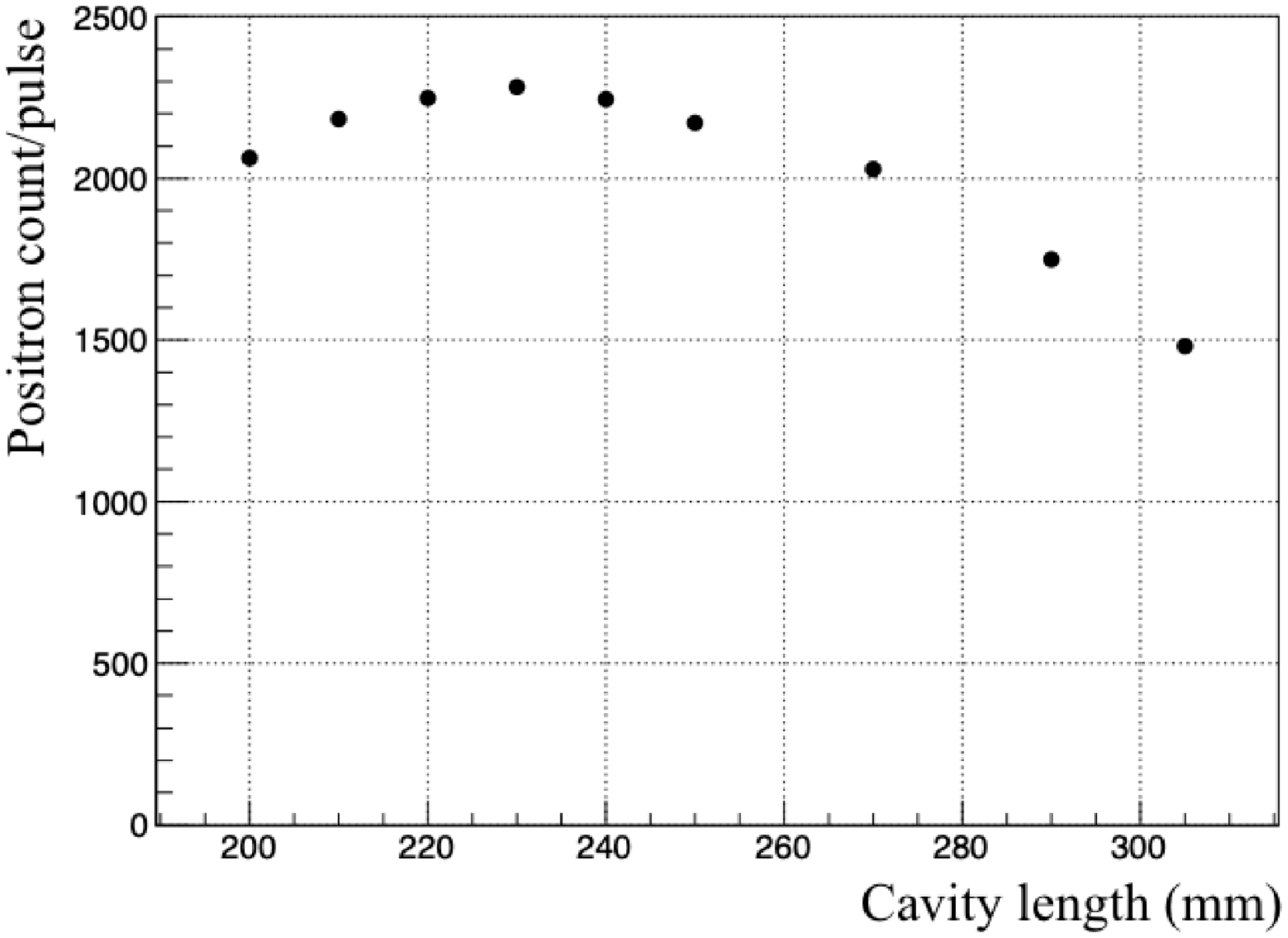}
     \caption{ Dependence of the number of positron detections on cavity length per a muon pulse ($10^6$ muons) calculated by GEANT4\cite{GEANT4_1}\cite{GEANT4_2}\cite{GEANT4_3}. The muon energy is $27.4\ \mathrm{MeV}/c$. A longer cavity accepts a larger number of muons in the cavity, but limits the solid angle of the decay positrons. The optimal length is 230 mm.}
     \label{zero-field-cavity-length-optimization.eps}
  \end{center}
  \end{figure}

As described in subsection \ref{subsection-cavity-structure}, decay positrons cannot pass through the sides. Thus, the solid angle of the decay positrons is limited in the case of the cavity for the zero-field experiment, which has a small diameter (Fig. \ref{axial-length-solid-angle}).
As illustrated in Fig. \ref{zero-field-cavity-length-optimization.eps}, the optimized cavity length for the measurement at a gas pressure of 1 bar is 230 mm.
\subsection{Structure of cavities and frequency charactaristics}\label{subsection-cavity-structure}
\begin{figure}[htbp]
  \begin{center}
     \includegraphics[width=\hsize,keepaspectratio]{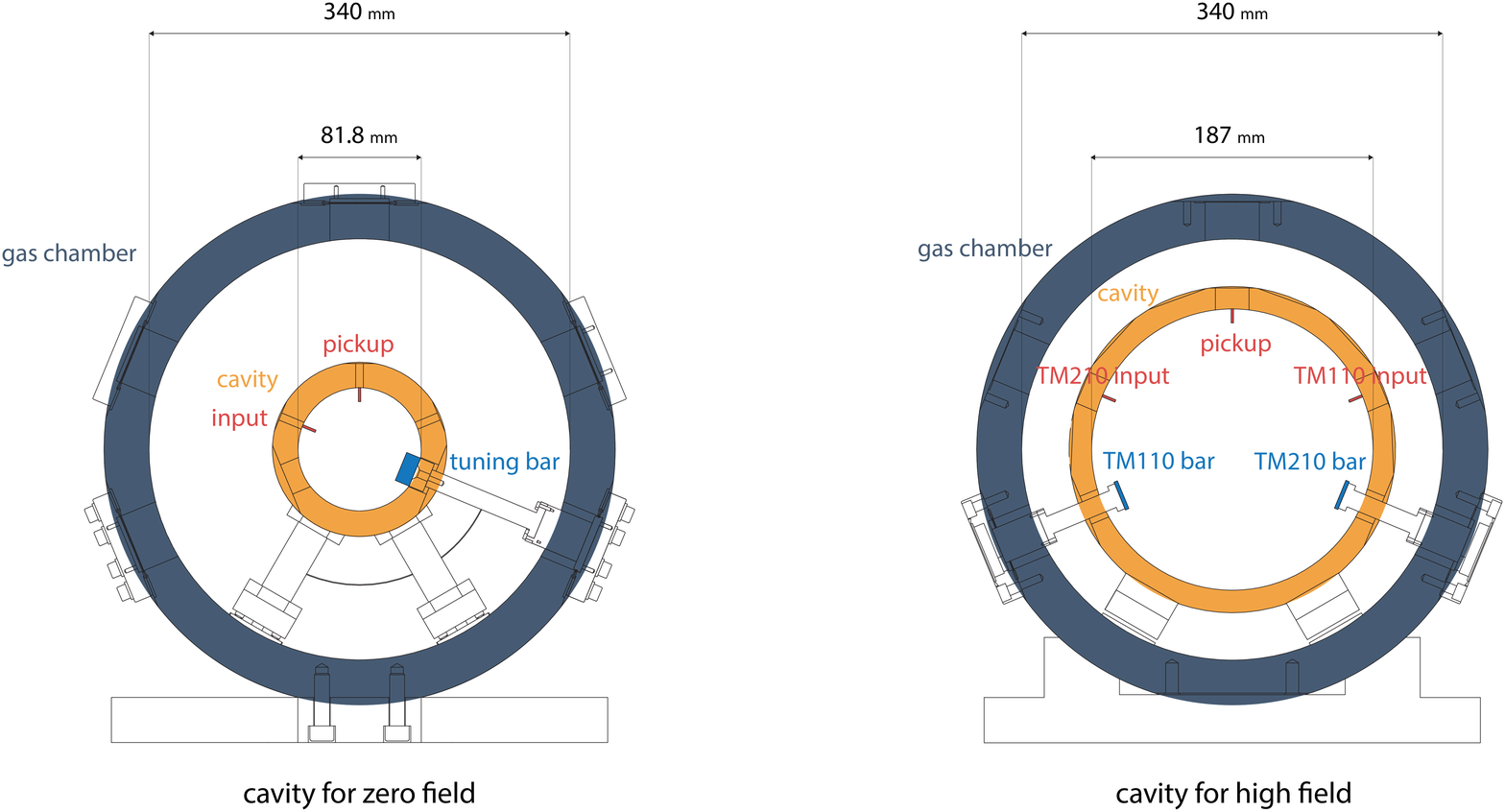}
     \caption{Cross-section views of cavities for zero-field and high-field.}
     \label{cavity-3dcad.eps}
  \end{center}
  \end{figure}

As mentioned in subsection \ref{subsection-cavity-length}, the cavity diameter for the zero-field experiment is 81.8 mm and the length is 230 mm. For the high-field experiment, the cavity diameter is 187 mm and the length is 304 mm (Fig. \ref{cavity-3dcad.eps}). The material of both cavities is oxygen-free copper.
 The front and back endcaps consist of 25 $\mathrm{\mu m}$ thickness copper foil, which the muons and decay positrons can penetrate.
 The cavity is located in the aluminum gas chamber that is filled with Kr gas.
  
  \begin{figure}[htbp]
   \begin{tabular}{cc}
     \begin{minipage}[t]{0.45\hsize}
           \centering
         \includegraphics[width=\hsize,keepaspectratio]{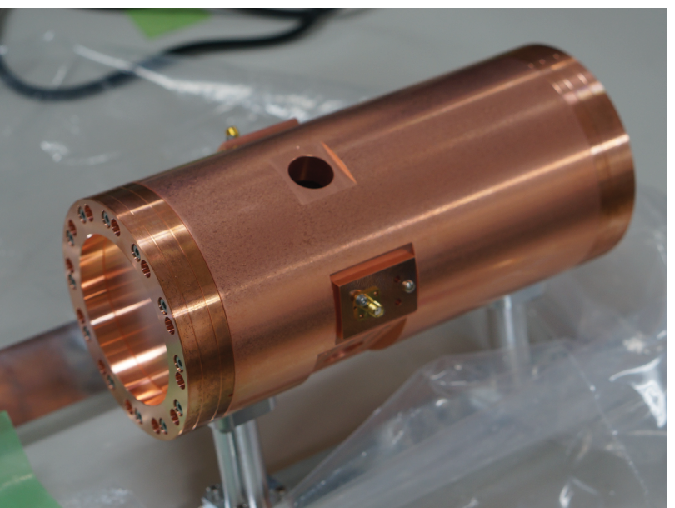}
         \caption{Microwave cavity for zero-field experiment.}
         \label{photo-zero-field-cavity.eps}

      \end{minipage} &
      \begin{minipage}[t]{0.45\hsize}
           \centering
         \includegraphics[width=\hsize,keepaspectratio]{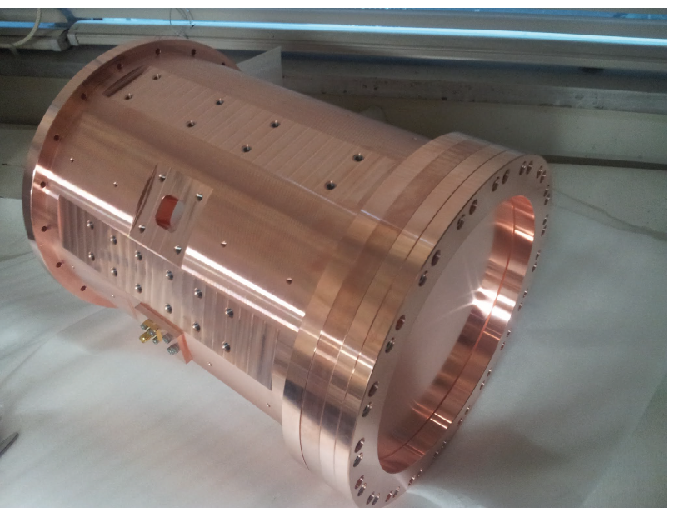}
         \caption{Microwave cavity for high-field experiment.}
         \label{photo-high-field-cavity.eps}
      \end{minipage}

   \end{tabular}
 \end{figure}

\par
The cavity for the zero-field experiment has one microwave input port and one output port to monitor the microwave power in the cavity.
The metal tuning bar is located on the opposite side of the input port. The microwaves from the signal generator (HP Agilent 8671B) are amplified by microwave amplifiers (Mini Circuits ZVE-8G) and transmitted to the cavity. The microwave signals are monitored by a power meter (R\&S NRP-Z51).
\par
\begin{figure}[htbp]
  \begin{center}
     \includegraphics[width=0.6\hsize,keepaspectratio]{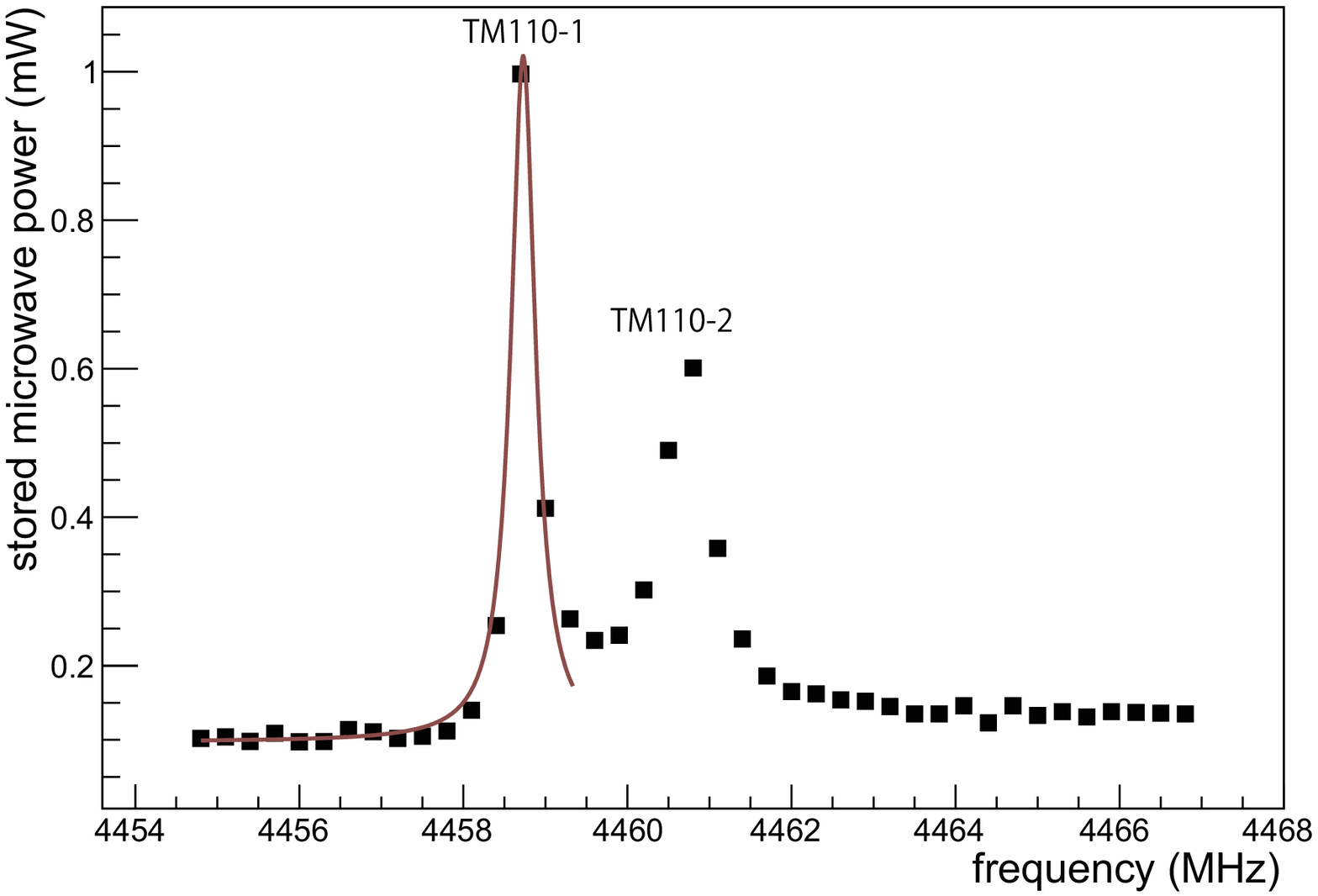}
     \caption{Measured Q-factor of TM110 mode. The Q factor is calculated from the FWHM of the resonance line shape and the FWHM is obtained from fitting with a Lorentz function.}
     \label{q-factor-tm110-zero-field}
  \end{center}
  \end{figure}

Figure \ref{q-factor-tm110-zero-field} depicts the frequency characteristics of the cavity for zero-field and experiment filled with air (the reflective index of air at 1 atm is $n \approx 1.000277$ \cite{Ciddor:96}), measured with a vector network analyzer. The two adjacent peaks are determined as the degenerated TM110 modes. The resonance frequency of the TM110 modes is 4.459 GHz. (Table \ref{table-frequency-characteristics-cavities}). As the TM110-1 mode is strongly coupled to the input port for TM110 and the TM110-2 mode is weakly coupled. According to the full width at half maximum (FWHM) of the TM110-1 peak, the Q factor is approximately $8.5 \times 10^3$.
\par
\begin{figure}[t]
  \begin{center}
  \includegraphics[width=0.7\hsize]{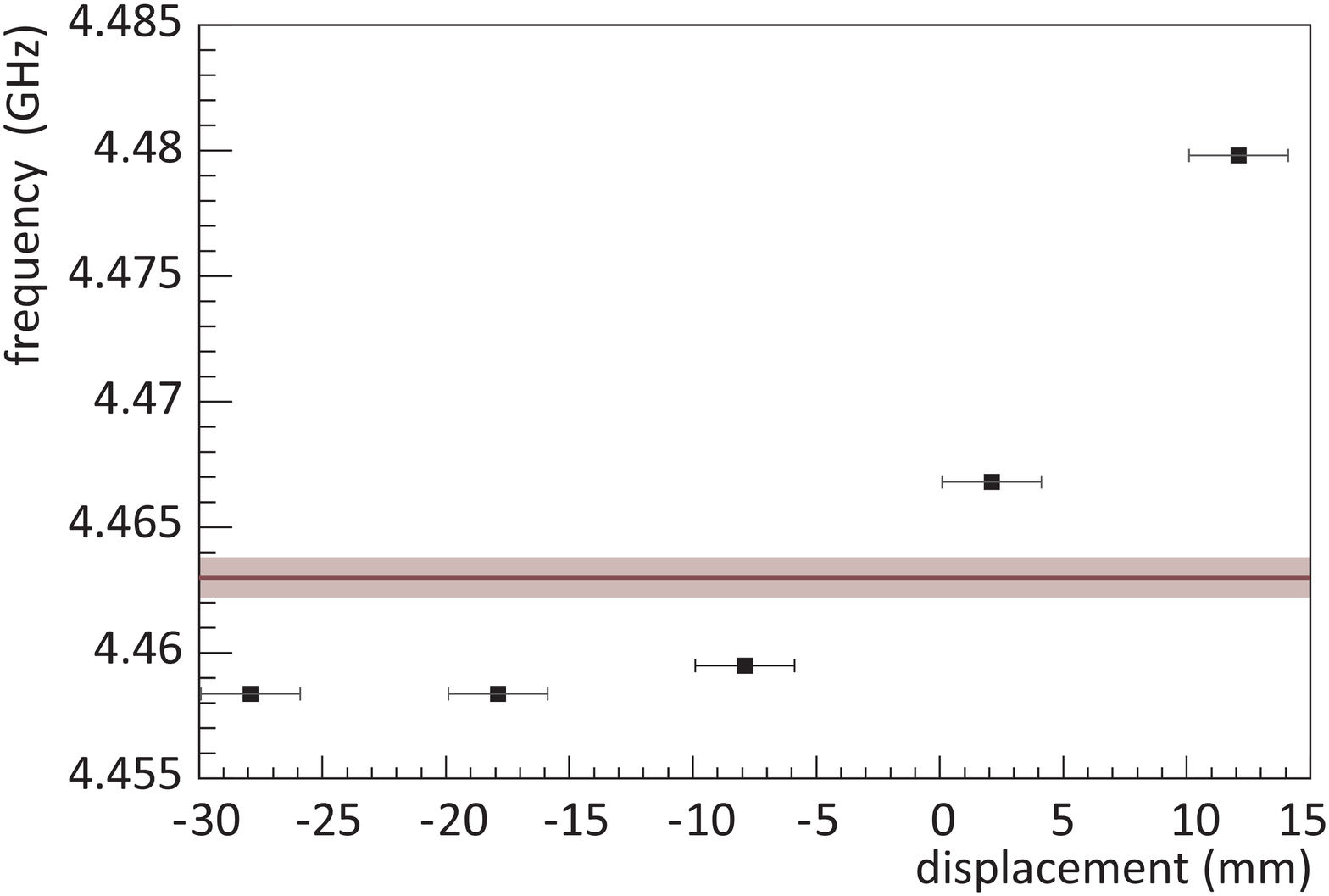}
  \caption{Relation between resonance frequency and displacement of tuning bar (square), transition frequency of HFS (red line), and sweep range to observe resonance line shape (red band). The insertion of the tuning bar made of conductive material is achieved by a rotary manipulator (IRC-70, IRIE-KOKEN), which has a step size of 4 $\mu$m.}
  \label{tuning-bar-performance-test-zero-field}
  \end{center}
  \end{figure}

The insertion of the cylindrical metal tuning bar with a diameter of 20 mm and length of 20 mm is achieved by a rotary manipulator (IRC-70, IRIE-KOKEN). The manipulator has a step size of 4 $\mu$m and it corresponds to a shift of resonance frequency of 4 kHz, which is sufficiently small compared to the natural linewidth of muonium ($145$ kHz).
Figure \ref{tuning-bar-performance-test-zero-field} presents the dependence of the resonance frequency on the displacement of the tuning bar (square), the transition frequency of the HFS (red line), and the sweep range to observe the
resonance line shape (red band). As indicated in Fig. \ref{tuning-bar-performance-test-zero-field}, the tunability of the cavity is approximately 20 MHz, which sufficiently wide to scan the entire resonance curve.
\par
The cavity for the zero-field experiment has both a microwave input port for TM110, and an output port to monitor the microwave power in the cavity.
The alumina tuning bars are located on the opposite side of each input port.
\par
\begin{figure}[t]
  \begin{center}
     \includegraphics[width=0.8\hsize,keepaspectratio]{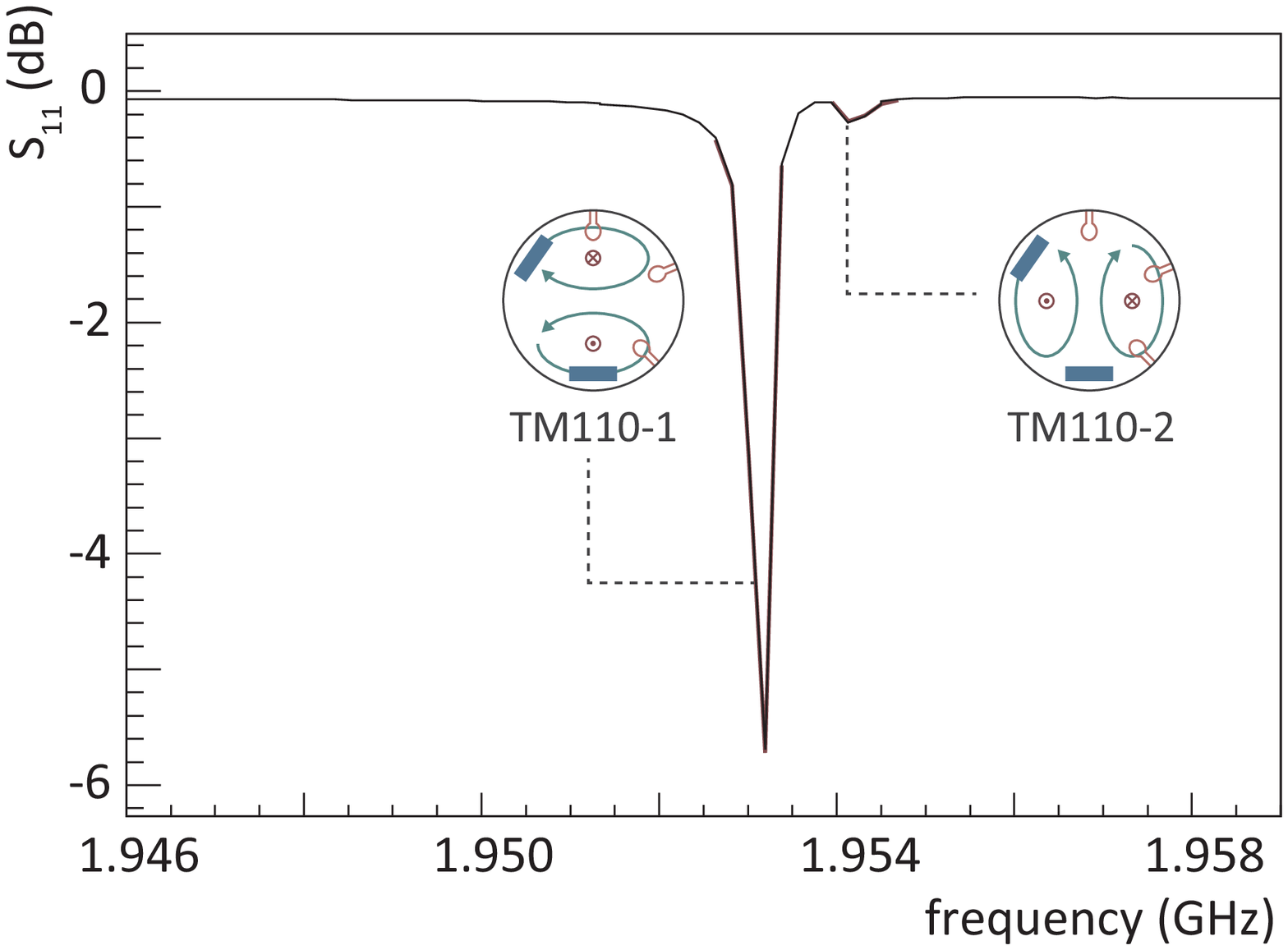}
     \caption{$S_{11}$ near TM110 modes from measurement. The two adjacent peaks are determined as the degenerated TM110 modes.}
     \label{TM110-without-tuningbar-compared-to-simulation-closeup-rev2}
  \end{center}
  \end{figure}
\begin{figure}[t]
  \begin{center}
     \includegraphics[width=0.8\hsize,keepaspectratio]{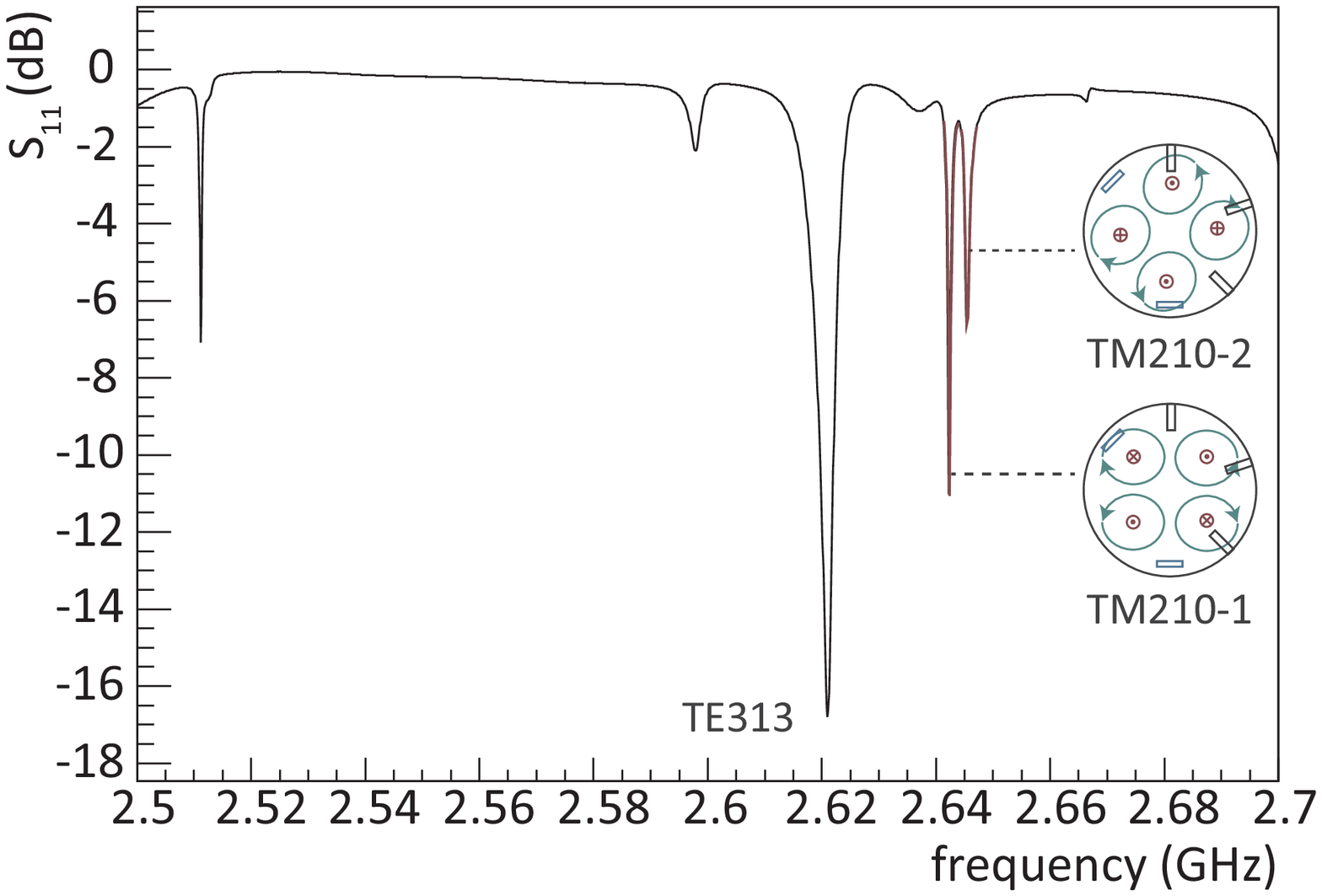}
     \caption{$S_{11}$ near TM210 modes from measurement. The two adjacent peaks are determined as the degenerated TM210 modes.}
     \label{TM210-without-tuningbar-compared-to-simulation-rev2}
  \end{center}
  \end{figure}

Figures \ref{TM110-without-tuningbar-compared-to-simulation-closeup-rev2} and \ref{TM210-without-tuningbar-compared-to-simulation-rev2} present the frequency characteristics of the cavity for the high-field experiment.
As with the cavity for the zero-field experiment, the TM110 mode is split into TM110-1 and TM110-2.
The Q factors calculated from the FWHM of the two peaks are $1.1\times 10^4$ for TM110-1 and $8.1\times 10^3$ for TM210-1.
\par
As the ratio of the cavity length to diameter is larger than that for the high-field experiment, the small size of the tuning bar affects the uniformity of the microwave field in the axial direction. Therefore, the tuning bar for the high-field experiment is long in the axial direction (200 mm $\times$ 20 mm $\times$ 2.5 mm).
The perturbation of the resonance frequency is approximated as \cite{pozar2005microwave}
\begin{equation}
\frac{\omega-\omega_{0}}{\omega_{0}} \approx - \frac{\int\int\int_{\Delta V}(\Delta\mu|H_{0}|^{2}+\Delta\epsilon|E_{0}|^{2})\mathrm{d}v}{\int\int\int_{V}(\mu|H_{0}|^{2}+\epsilon|E_{0}|^{2})\mathrm{d}v} \
\label{equation-cavity-perturbations},
\end{equation}
where $\epsilon$ and $\mu$ are vacuum permittivity and permeability, $\Delta\epsilon$ and $\Delta\mu$ are permittivity and permeability in the tuning bar,  $\omega_{0}$ is the resonance frequency of the original cavity and $\omega$ is that of the perturbed cavity. $E_{0}$ and $H_{0}$ are the original electric and magnetic fields, respectively.
\par
\begin{figure}[tbp]
  \begin{center}
     \includegraphics[width=\hsize,keepaspectratio]{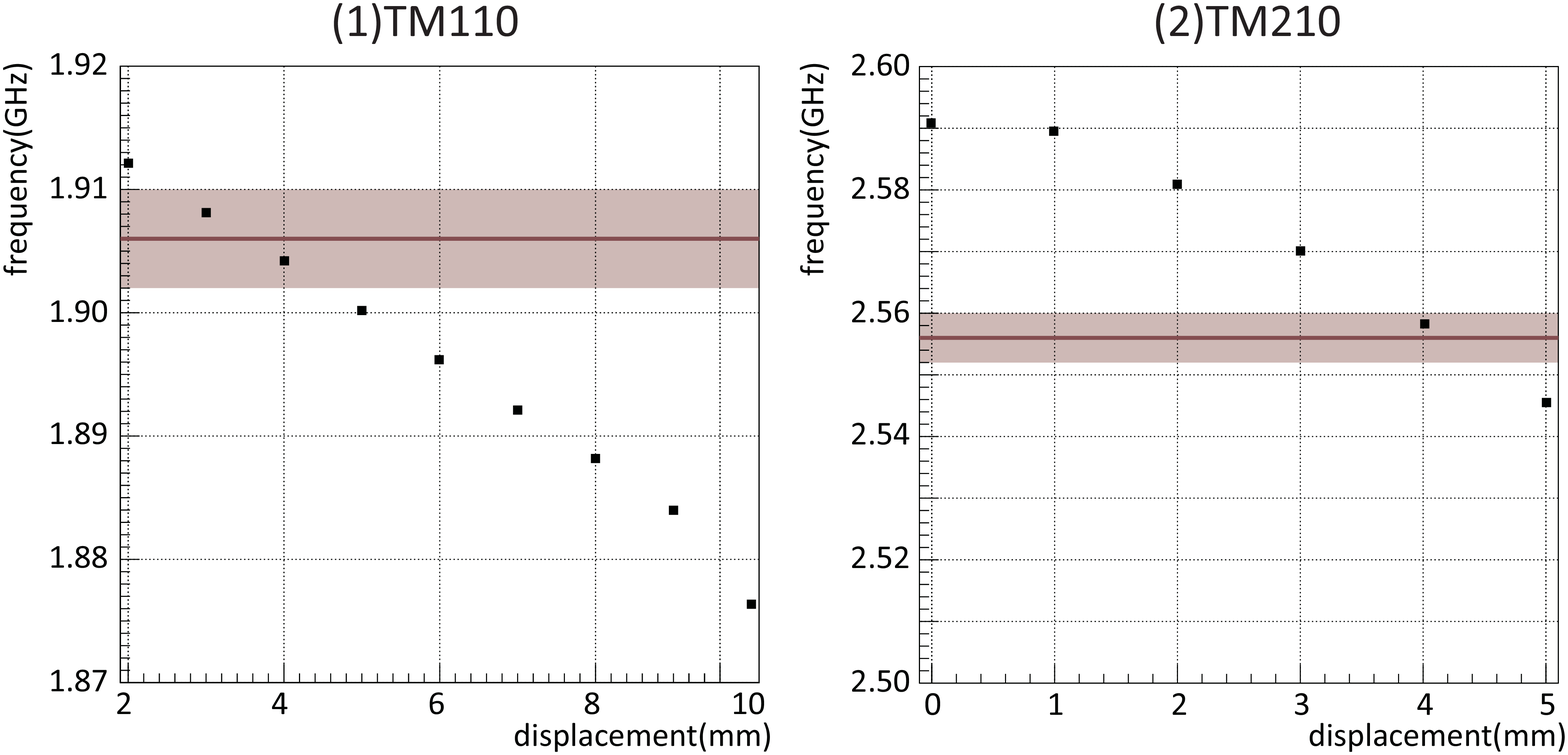}
     \caption{Relation between resonance frequencies of (1) TM110, (2) TM210 and displacement of tuning bars (square), transition frequency of HFS (red line), and sweep range to observe resonance line shape (red band).}
     \label{tm110-tm210-tuning}
  \end{center}
  \end{figure}

Figure \ref{tm110-tm210-tuning} displays the measured resonance frequencies of TM110-1 and TM210-1.
The tuning bar is inserted into the cavity using a piezo positioner (ANPz101eXT12/RES). The tunable range is 12 mm and the minimum step size is 50 nm, which correspond to a sweeping range of 36 MHz and frequency perturbation of 150 Hz in the case of the TM110 mode. These are sufficient to cover the linewidth of the muonium transition resonance; that is, hundreds of kHz.

\begin{table}[htb]
   \caption{Measured frequency characteristics of microwave cavities.}

   \begin{center}
   \begin{tabular}{|l||c|c|c|}
   \hline
   Cavity & For zero-field & \multicolumn{2}{c|}{For high-field} \\
   \hline
   \hline
   Diameter & 81.8 mm & \multicolumn{2}{c|}{187 mm}  \\
   Length & 230 mm & \multicolumn{2}{c|}{304 mm}  \\
  \hline
   Resonance mode & TM110 & TM110 & TM210  \\
   \hline
   Resonance frequency & 4.459 GHz & 1.953 GHz& 2.645 GHz\\
   Q factor & $8.5 \times 10^3$& $1.1 \times 10^4$ & $8.1 \times 10^3$\\
   \hline
   \end{tabular}
   \end{center}

   \label{table-frequency-characteristics-cavities}
   \end{table}
\section{Systematic uncertainties}
Past experiments estimate systematic uncertainties with a theoretical microwave distribution that ignore the effect of tuning bars or microwave antennas which is now important for our high precision experiment. To evaluate systematic uncertainties from such a fluctuation of microwave fields and confirm the requirement of the microwave system, we use a microwave field distribution calculated from the finite element method.

\begin{figure}[htbp]
\begin{center}
   \includegraphics[width=\hsize,keepaspectratio]{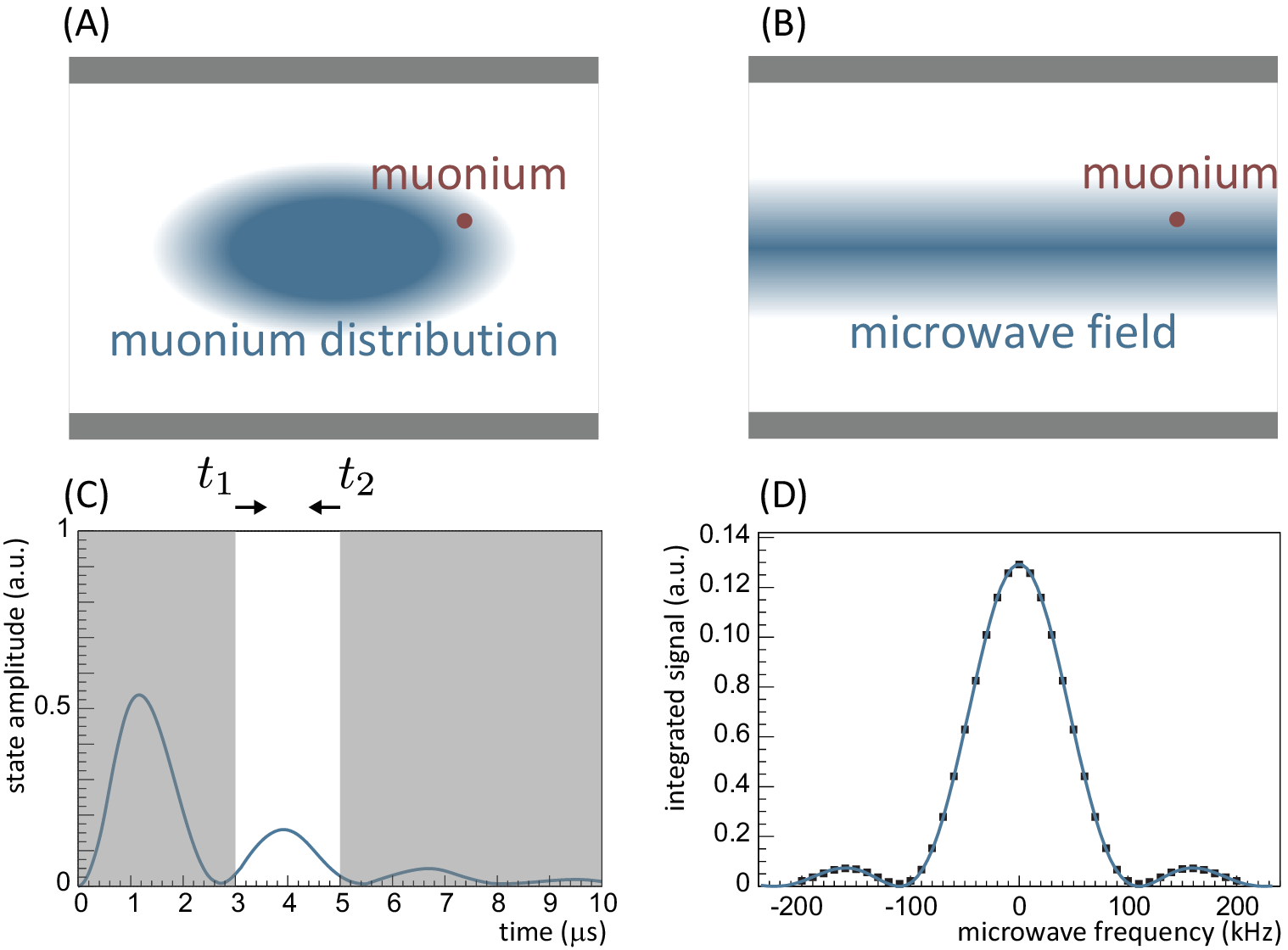}
   \caption{Procedure of evaluation of systematic uncertainties with microwave field distribution calculated from the finite element method. (A) Muonium distribution is obtained by the Geant4 assuming a certain pressure of Kr gas. One of the muoniums is picked in the distribution. (B) Obtain the perpendicular component of the microwave magnetic field at the muonium's position from the result of MWS calculation. (C) Calculate the time evolution of the excited state amplitude with a perpendicular component of the microwave magnetic field and the difference between transition and resonance frequencies. The transition probability is calculated by integrating the time evolutions from $t_{1}$ to $t_{2}$. (D) The resonance line is formed by integrating the transition probabilities of all muoniums during a microwave resonance scan. 
   }
   \label{simulation-scheme.eps}
\end{center}
\end{figure}

\begin{figure}[htbp]
   \begin{center}
      \includegraphics[width=\hsize,keepaspectratio]{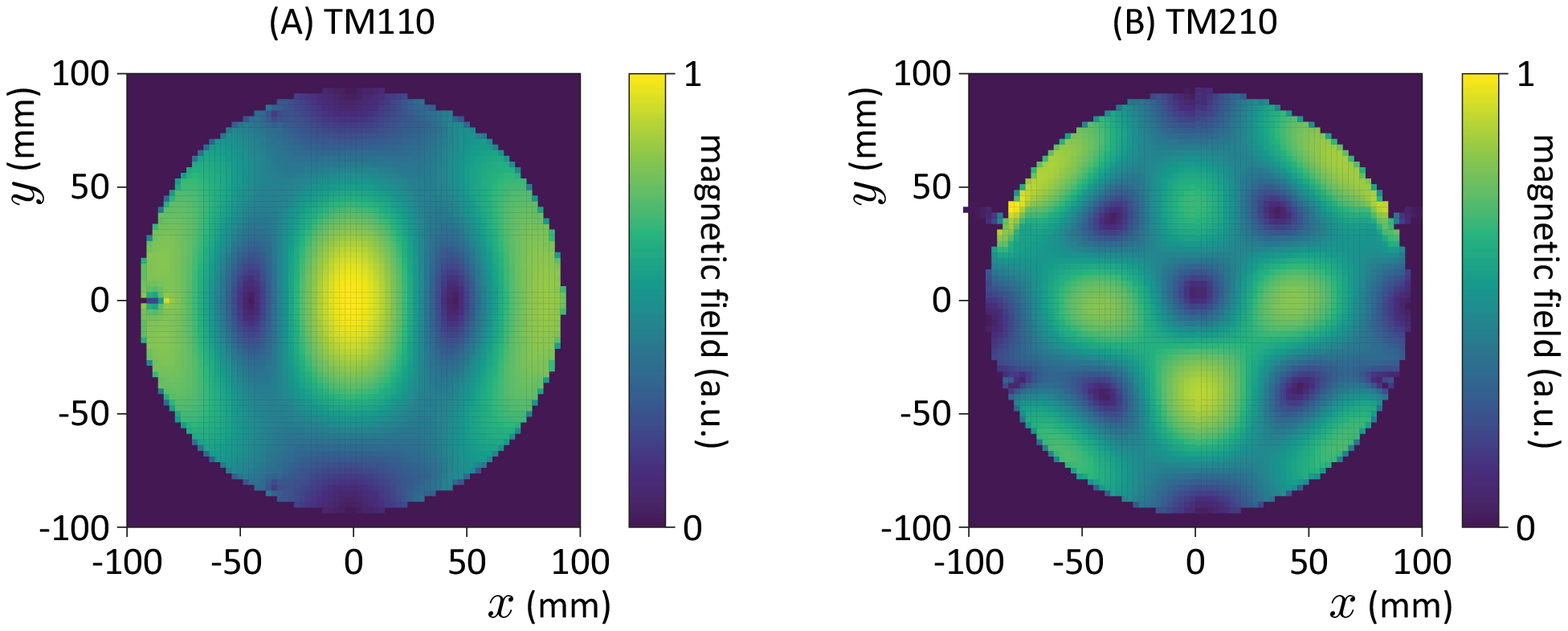}
      \caption{Cross-sectional view at the center of the cavity of the magnetic field strength of microwave perpendicular to the cavity axis . The distribution is calculated by MWS assuming that applying (A) TM110 resonance and (B) TM210 resonance to the cavity for high field.}
      \label{rf-b-map-xy-0mm.eps}
   \end{center}
   \end{figure}

Figure \ref{simulation-scheme.eps} shows the procedure of evaluation of systematic uncertainties with microwave field distribution calculated from the MWS. Muonium distribution is obtained by the Geant4 assuming a certain pressure of Kr gas. One of the muoniums is picked in the distribution and the microwave magnetic field at the muonium's position is obtained from the result of the MWS calculation (Fig. \ref{rf-b-map-xy-0mm.eps}). Since the MWS calculation is using a finite element analysis with voxel mesh, quadratic interpolation is used to calculate the field strength at the point. Then the time evolution of the excited state amplitude is obtained with the perpendicular component of the microwave magnetic field and the difference between transition and resonance frequencies. The transition probability is calculated by integrating the time evolutions in a certain time window. The resonance line is formed by integrating the transition probabilities of all muoniums with each microwave frequencies.
Thus the systematic uncertainties from the microwave field can be evaluated by calculating the resonance line with this process to evaluate the uncertainty of the center value of the resonance peak.

We use the microwave field obtained by MWS and the typical muonium distribution to estimate systematic uncertainties from the time fluctuation of the microwave field and the variation by the position of tuning bars.

\section{Systematic uncertainties from microwave power drift and fluctuation of microwave
field}\label{section-systematic-unvertainties-RF}

\begin{figure}[htbp]
   \begin{center}
      \includegraphics[width=\hsize,keepaspectratio]{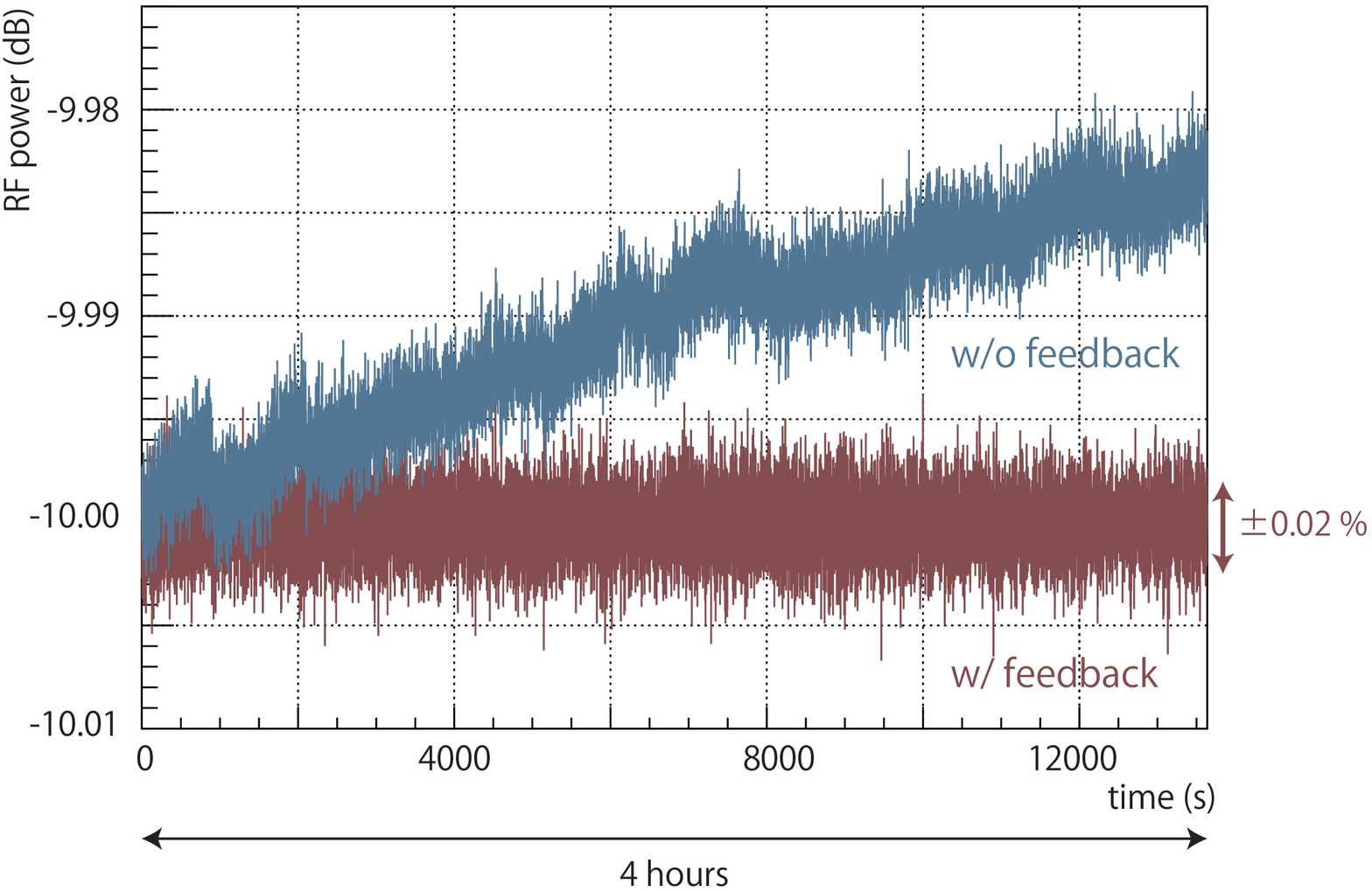}
      \caption{Time fluctuation of microwave power from the signal generator without the feedback from the power meter (blue line) or with the feedback (red line). The microwave power drift is suppressed by the feedback system and stabilized within 0.02 \%.
      }
      \label{rf-power-feedback}
   \end{center}
   \end{figure}

One of the main source of a systematic uncertainty from microwave power is power drift of signal generator which is enabled to suppressed by the feedback system and stabilized within 0.02 \% (Fig. \ref{rf-power-feedback}). If we assume that no correlation exists between the microwave power and frequency, no systematic uncertainty arises and the line shape is simply broadened, which is negligible compared to the statistical uncertainty.

\begin{figure}[htbp]
\begin{center}
   \includegraphics[width=\hsize,keepaspectratio]{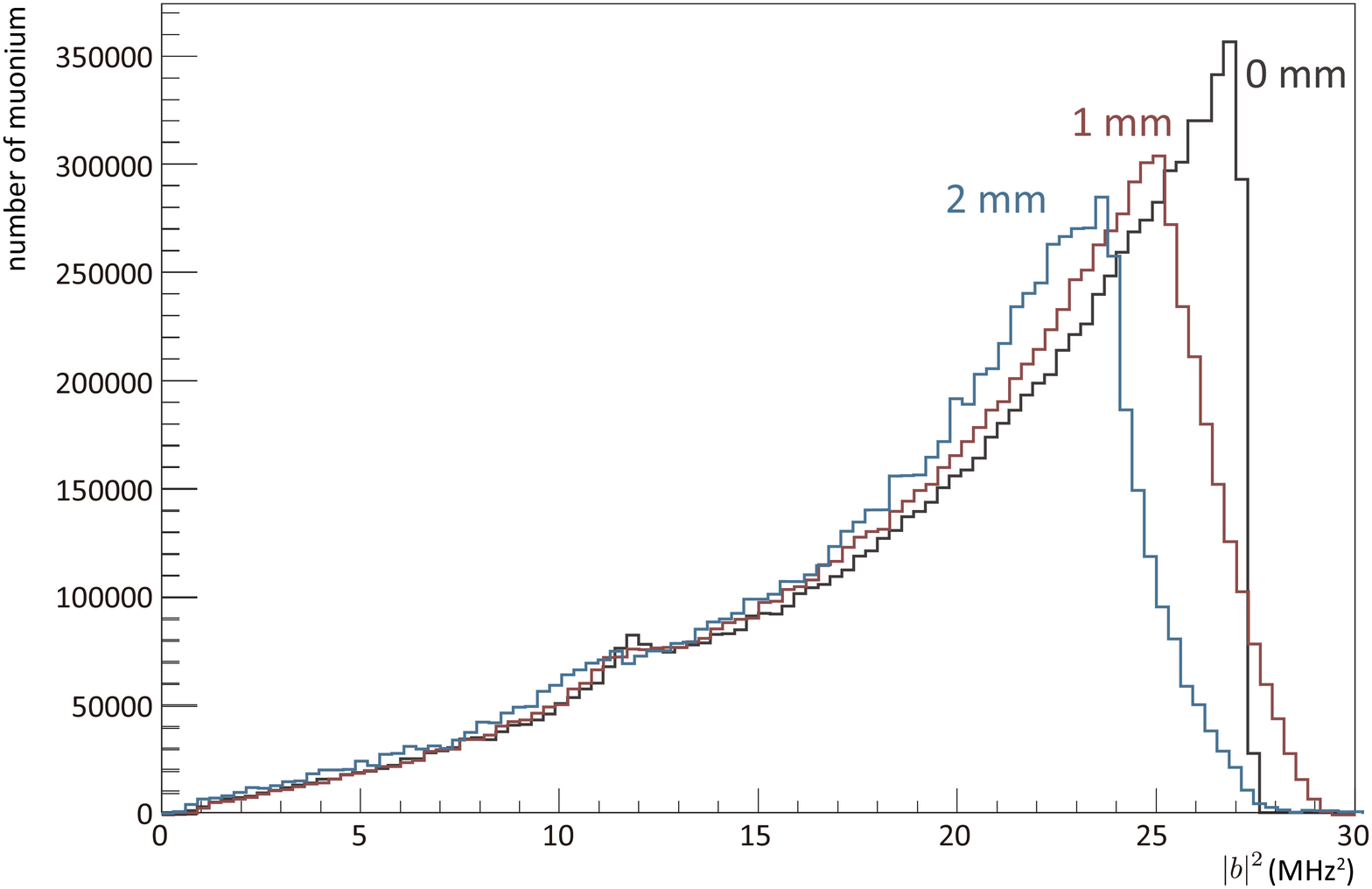}
   \caption{Histogram of the strength of magnetic field of microwave perpendicular to the cavity axis ($|b|^{2}$) when tuning bar displacement is 0, 1 and 2 mm. It assumed the TM110 resonance mode at the high-field cavity. the number of muons is $5.0 \times 10^{6}$ and the microwave fields were calculated by MWS.}\label{mu-hfs-b-distribution-map}
\end{center}
\end{figure}

The microwave field in the cavity is affected by the displacement of the tuning bars. Figure \ref{mu-hfs-b-distribution-map} shows a histogram of the microwave power $|b|^{2}$ affecting each muon in the cavity in the case of tuning bar displacements of 0, 1 and 2 mm assuming the TM110 resonance mode at the high-field cavity (1 mm of displacement corresponds to a frequency shift of 8 MHz). In this case, $b$ is defined by the oscillation frequency $\Gamma$ of two states ($\ket{2}$ to $\ket{1}$ or $\ket{4}$ to $\ket{3}$) as follows:
\begin{equation}
\Gamma = \sqrt{\Delta\nu^2 +4|b|^2},
\end{equation}
where $\Delta\nu$ is the frequency difference between the transition frequency and resonance frequency of the cavity \cite{PhysRevA.3.871}\cite{PhysRevA.5.2338}. 
$b$ is a Rabi frequency between state $i$ to state $j$ defined as 
\begin{equation}
 b = \frac{1}{2\hbar}  <i |(g'_{e}\mu_{e}^{B} -g'_{\mu} \mu^{B}_{\mu})B_{1} |j>,
\end{equation}
where $B_{1}$ is a microwave field strength. 
The number of muons is $5.0 \times 10^{6}$ and the microwave fields were calculated by MWS. As the peak position of both the microwave field and muonium distribution is located on the center axis of the cavity, this histogram is almost monotonically increasing.
\par
The transition frequency is obtained by fitting the resonance function weighted by the histogram of the perpendicular components of the microwave magnetic field strength.
If the distribution of the microwave in the cavity is not distorted by the tuning bar, the resonance line is the sum of the resonance lineshape from each muon with different perpendicular components of the microwave magnetic field strength. 
On the other hand, if the distribution of the microwave in the cavity is distorted by the tuning bar as shown in Fig. \ref{mu-hfs-b-distribution-map}, the histogram of the magnetic field strength fluctuates during a frequency scan.
In case the tuning bar displacement is 0 and 50 $\mu$m, which corresponds to about 400 kHz frequency scan, the fitting error is about 9 Hz if we use a fitting function that assumes the static histogram of the magnetic field strength.
However, if the fitting function assumes a histogram to change linearly from 0 mm to 50 $\mu$m, the uncertainties were calculated as shown in Table \ref{systematic-uncertainty-from-rf}. Both uncertainties of $\nu_{\mathrm{HFS}}$ from microwave power and displacement of tuning bars are 
enough for achieving a precision of several ppb in the measurement.
\begin{table}[htb]
   \caption{Systematic uncertainty from microwave field fluctuation}
\begin{center}
\begin{tabular}{|l||l|l|}
\hline
 & Microwave power & Tuning bars\\
\hline
\hline
$\nu_{12}$& 2 Hz& 1 Hz\\
$\nu_{34}$& 2 Hz& 2 Hz\\
$\nu_{\mathrm{HFS}}$ in high-field& 3 Hz (0.7 ppb)& 3 Hz (0.7 ppb)\\
$\nu_{\mathrm{HFS}}$ in zero-field& 3 Hz (0.7 ppb)& 2 Hz (0.5 ppb)\\
$\mu_{\mu}/\mu_{p}$& 9 ppb& 7 ppb\\
\hline
\end{tabular}
\end{center}

\label{systematic-uncertainty-from-rf}
\end{table}
\section{Conclusions}
Microwave cavities have been prepared for Mu-HFS spectroscopy experiments in both the zero-field and high-field that are planned at J-PARC.
We developed a longer microwave cavity for the high-field which is compared to the previous LAMPF experiment to enable measurement at a gas pressure of 0.3 bar or below, thereby taking advantage of the larger uniform region of the magnetic field.
The frequency characteristics of both cavities were evaluated and they fulfilled the requirements for tunability of the resonance frequencies.
\par
We use a microwave field distribution calculated from the finite element method which is not considered in the LAMPF measurement to evaluated systematic uncertainties. Systematic uncertainties from such a fluctuation of microwave power is are ppb for high and zero field experiment, and from  displacement of tuning bars are 0.7 ppb for high field experiment and 0.5 ppb for zero field experiment. 
We aim to achieve a precision of several ppb in the measurement and both uncertainties from microwave power and displacement of tuning bars are fulfilled this requirement.
\section{Acknowledgment}
The MuSEUM experiment has been proposed as a S-type user program proposal number 2011MS01. The authors wish to acknowledge the help and expertise of the staff of J-PARC MLF. This research was supported by JSPS KAKENHI, Grant Numbers 23244046 and 26247046.
\par
\bibliographystyle{ptephy}
\bibliography{PTEP-RF-MuSEUM}
\end{document}